\documentclass{article}
\usepackage{spconf}
\ninept %
\usepackage{multirow, graphicx}
\usepackage{times}            %
\usepackage[english]{babel}   %
\usepackage[ansinew]{inputenc}%
\usepackage[T1]{fontenc}      %
\usepackage[sort&compress,numbers]{natbib}	%
\usepackage{amsmath,amssymb}
\usepackage{graphicx}
\usepackage[colorlinks=false,pdfborder={0 0 0}]{hyperref}
\usepackage{units}

\usepackage{mathtools}
\newcommand{\norm}[1]{\left\lVert#1\right\rVert}
\usepackage{xfrac}

\newcommand{\D}[1]{\mathrm{d}{#1}}

\newcommand{\bX}{\mathbf X}
\newcommand{\btX}{\widetilde{\mathbf X}}
\newcommand{\bY}{\mathbf Y}

\newcommand{\bS}{\mathbf S}
\newcommand{\teps}{t_{\epsilon}}
\newcommand{\trsp}{t_{\text{rsp}}}

\usepackage{comment}

\DeclareMathOperator*{\argmin}{arg\,min}

\usepackage[font=small,skip=0pt]{caption}
\usepackage[nolist, nohyperlinks]{acronym}
\begin{acronym}
\acro{bb}[BB]{Brownian bridge}
\acro{sgm}[SGM]{score-based generative model}
\acro{snr}[SNR]{signal-to-noise ratio}
\acro{gan}[GAN]{generative adversarial network}
\acro{vae}[VAE]{variational autoencoder}
\acro{ddpm}[DDPM]{denoising diffusion probabilistic model}
\acro{stft}[STFT]{short-time Fourier transform}
\acro{istft}[iSTFT]{inverse short-time Fourier transform}
\acro{sde}[SDE]{stochastic differential equation}
\acro{ode}[ODE]{ordinary differential equation}
\acro{ou}[OU]{Ornstein-Uhlenbeck}
\acro{pc}[PC]{Predictor-Corrector}
\acro{ve}[VE]{Variance Exploding}
\acro{dnn}[DNN]{deep neural network}
\acro{pesq}[PESQ]{Perceptual Evaluation of Speech Quality}
\acro{se}[SE]{speech enhancement}
\acro{tf}[T-F]{time-frequency}
\acro{elbo}[ELBO]{evidence lower bound}
\acro{WPE}{weighted prediction error}
\acro{PSD}{power spectral density}
\acro{RIR}{room impulse response}
\acro{SNR}{signal-to-noise ratio}
\acro{LSTM}{long short-term memory}
\acro{POLQA}{Perceptual Objectve Listening Quality Analysis}
\acro{SDR}{signal-to-distortion ratio}
\acro{ESTOI}{Extended Short-Term Objective Intelligibility}
\acro{ELR}{early-to-late reverberation ratio}
\acro{TCN}{temporal convolutional network}
\acro{DRR}{direct-to-reverberant ratio}
\acro{nfe}[NFEs]{number of function evaluations}
\acro{rtf}[RTF]{real-time factor}
\acro{em}[EuM]{Euler-Maruyama}
\acro{ald}[ALD]{annealed Langevin Dynamics}
\acro{crp}[CRP]{correcting the reverse process}
\acro{dsm}[DSM]{denoising score matching}
\end{acronym}

\title{Single and Few-step Diffusion for Generative Speech Enhancement}

\name{Bunlong Lay, Jean-Marie Lermercier, Julius Richter\thanks{This work has been funded by the German Research Foundation (DFG) in the transregio project Crossmodal Learning (TRR 169).}, Timo Gerkmann}
\address{Signal Processing, University Hamburg, Hamburg, Germany \\bunlong.lay@uni-hamburg.de}

\begin{document}

\maketitle

\begin{abstract}
Diffusion models have shown promising results in single-channel speech enhancement, using a task-adapted diffusion process for the conditional generation of clean speech given a noisy mixture. However, at test time, the neural network used for score estimation is called multiple times to solve the iterative reverse process. This results in a slow inference process and causes discretization errors that accumulate over the sampling trajectory. 
In this paper, we address these limitations through a two-stage training approach. In the first stage, we train the diffusion model the usual way using the generative denoising score matching loss. In the second stage, we compute the enhanced signal by solving the reverse process and compare the resulting estimate to the clean speech target using a predictive loss. We show that using this second training stage enables achieving the same performance as the baseline model using only 5 function evaluations instead of 60 function evaluations. 
While the performance of usual generative diffusion algorithms drops dramatically when lowering the number of function evaluations to obtain single-step diffusion, we show that our proposed method keeps a steady performance and therefore largely outperforms the diffusion baseline in this setting and also generalizes better than its predictive counterpart\footnote{Find code and audio examples: \href{https://github.com/sp-uhh/sgmse_crp}{https://github.com/sp-uhh/sgmse\_crp}}. 
\end{abstract}

\section{Introduction} \label{sec:intro}

The objective of speech enhancement (SE) is to retrieve the original clean speech signal from a noisy mixture that is affected by environmental noise ~\cite{hendriks2013dft}. Traditional approaches attempt to leverage statistical relationships between the clean speech signal and the surrounding environmental noise  ~\cite{gerkmann2018book_chapter}. In recent years, various machine learning techniques have been introduced, treating SE as a discriminative learning task \cite{wang2018supervised, luo2019conv} resulting in so-called \emph{predictive} approaches.

In contrast to predictive approaches that learn a direct mapping from noisy to clean speech, generative approaches learn a prior distribution over clean speech data. 
Recently, so-called \emph{score-based generative models}, also known as \emph{diffusion models} are introduced for SE \cite{lu2021study, lu2022conditional, paper, journal}.
The fundamental concept behind these models involves iteratively adding Gaussian noise to the data using a discrete and fixed Markov chain known as the \emph{forward process}, thereby transforming data into a tractable distribution such as the standard normal distribution. Then, a neural network is trained to invert this diffusion process in a so-called \emph{reverse process}~\cite{ho2020denoising}.
By letting the step size between two discrete Markov chain states approach zero, the discrete Markov chain becomes a continuous-time stochastic process that can be described using \acp{sde}. This offers more flexibility and opportunities than approaches based on discrete Markov chains \cite{song2021sde}. For example, it allows the use of general-purpose \ac{sde} solvers to numerically integrate the reverse process, which can affect the performance and number of iteration steps. 

Solving the reverse process, also called reverse SDE when employing SDEs, means enhancing noisy mixtures  as the reverse SDE transforms the distribution of noisy speech into the distribution of clean speech. To solve the reverse SDE at inference, it is essential to evaluate the score function of the perturbed data, which is typically intractable. Thus, a neural network, also called \emph{score model}, is trained to approximate the required score function. The score model is typically trained using the \ac{dsm}  loss \cite{vincent2011connection}, which has shown great success for SE \cite{journal, lemercier2022storm, lay202interspeech}. However, solving the reverse SDE differs from the training condition in various aspects. 
First, there exists the so-called \textit{prior mismatch} \cite{lay202interspeech}, which we define as the difference between the terminating distribution of the forward process and the initial distribution used to solve the reverse SDE. 
Second, the reverse process introduces a discretization error when using a numerical solver. This error can be mitigated by using more sophisticated solvers, at the cost of increased \ac{nfe}. 
Last, the score model itself is not a perfect estimator of the score function and produces errors.  As the aforementioned errors accumulate over the reverse trajectory, the obtained diffusion states can deviate from the distribution seen during training, which further impairs the score model performance.

This work aims to mitigate these errors by employing a two-stage training. We first train on the \ac{dsm} objective. In the second stage we fine-tune the score model with a predictive loss to correct the discretization and score model errors. To do so, we pick a reverse diffusion solver, e.g. the \ac{em} first-order method in our case, to obtain a clean speech estimate. Then, we fine-tune the score model parameters with respect to the mean squared error (MSE) loss 
between the estimate and the clean speech data. Backpropagating through the whole reverse trajectory is computationally infeasible when using many steps. Therefore, we propose here to only use the gradients accumulated during the last reverse diffusion step to update the parameters. %

We compare our approach against (a) our generative baseline BBED \cite{lay202interspeech}, (b) a predictive baseline using a similar network architecture as used for score estimation, as well as (c) the StoRM approach \cite{lemercier2022storm}, which also combines a generative loss and a predictive loss. 
We show that the proposed method is very robust to a decrease in the NFEs used for reverse diffusion as opposed to the purely generative approach and StoRM \cite{lemercier2022storm}. In fact, with only 5 NFEs the proposed method performs virtually the same as the generative baseline with 60 NFEs. 
Furthermore, when taking only one reverse diffusion step, we achieve competitive SE performance and we show that the proposed generative approach generalizes better to unseen data than the predictive baseline. 

\section{Background}
\subsection{Notation and data representation} \label{sec:representation}
The task of SE is to estimate the clean speech signal $\mathbf S$ from a noisy mixture $\mathbf Y = \mathbf S+\mathbf N$, where $\mathbf N$ is environmental noise. 
All variables in bold are the coefficients of a compressed complex-valued spectrogram obtained through \ac{stft} and magnitude-compression, e.g. $\mathbf Y \in \mathbb{C}^d$ and $d=KF$ with $K$ number of \ac{stft} frames and $F$ number of frequency bins. As in \cite{journal}, we use the magnitude compression $\bar{c} = \beta |c|^\alpha \mathrm e^{i \angle(c)}$
with parameters $\beta, \alpha > 0$ to compensate for the typically heavy-tailed distribution of \ac{stft} speech magnitudes~\cite{gerkmann2010empirical}. We set $\beta=0.15, \alpha=0.5$ as in previous work \cite{journal, lay202interspeech, lemercier2022storm}. 

\subsection{Stochastic Differential Equations} \label{sec:sde}
Following \cite{paper, journal, lay202interspeech}, we model the forward process of the score-based generative model with an \ac{sde} defined on $0 \leq t < T_{\text{max}}$:
\begin{equation} \label{eq:fsde}
    \D{\mathbf \bX_t} =
       \mathbf f(\mathbf \bX_t, \mathbf \bY) \D{t}
        + g(t)\D{{\mathbf w}},
\end{equation}
where $\mathbf w$ is the standard Wiener process \cite{kara_and_shreve}, $\mathbf \bX_t$ is the current process state with initial condition $\mathbf \bX_0 = \mathbf \bS$, and $t$ a continuous diffusion time variable describing the progress of the process ending at the last diffusion time $T_{\text{max}}$.
The drift term $\mathbf f(\mathbf \bX_t, \mathbf \bY) \D{t}$ can be integrated by Lebesgue integration \cite{rudin}, and the diffusion term $g(t)\D{{\mathbf w}}$ follows Ito integration \cite{kara_and_shreve}. 
The diffusion coefficient $g$ regulates the amount of Gaussian noise that is added to the process, and the drift $\mathbf f$ affects the mean and the variance of $\mathbf \bX_t$ in the case of linear SDEs (see \cite[Eq. (6.10), (6.11)]{kara_and_shreve}). The process state $\mathbf \bX_t$ follows a Gaussian distribution \cite[Section 5]{kara_and_shreve} called \emph{perturbation kernel}:
\begin{equation}
\label{eq:perturbation-kernel}
    p_{0t}(\mathbf \bX_t|\mathbf \bX_0, \mathbf \bY) = \mathcal{N}_\mathbb{C}\left(\mathbf \bX_t; \boldsymbol \mu(\mathbf \bX_0, \mathbf \bY, t), \sigma(t)^2 \mathbf{I}\right).
\end{equation}
Under mild regularity conditions, each forward SDE \eqref{eq:fsde} can be associated to a reverse SDE \cite{anderson1982reverse}:
\begin{equation}\label{eq:plug-in-reverse-sde}
    \D{\mathbf \btX_t} =
        \left[
            -\mathbf f(\mathbf \btX_t, \mathbf \bY) + g(t)^2\mathbf  \nabla_{\mathbf \btX_t} \log p_t(\mathbf \btX_t|\mathbf \bY)
        \right] \D{t}
        + g(t)\D{\widetilde{\mathbf w}}\,,
\end{equation}
where
$\D{\widetilde{\mathbf w}}$ is a Wiener process going backward through the diffusion time. In particular, the reverse process starts at $t=T$ and ends at $t=0$. Here $T < T_{\text{max}}$ is a parameter that needs to be set for practical reasons, as the last diffusion time $T_{\text{max}}$ is only reached in limit.
The \emph{score function} $\nabla_{\mathbf X_t} \log p_t(\mathbf \bX_t|\mathbf \bY)$ is approximated by a neural network called \emph{score model} $\mathbf s_\theta(\mathbf \bX_t, \mathbf \bY, t)$, which is parameterized by a set of parameters $\theta$.
Assuming $\mathbf s_\theta$ is available, we generate an estimate of clean speech data from $\mathbf \bY$ by solving the reverse SDE.

\subsection{Solving the reverse process} \label{sec:solve_rsde}

To solve the reverse SDE the first-order EuM method can be employed to keep the \ac{nfe} low. For the \ac{em} method, a discretization schedule $(t_{N}, t_{N-1}, \dots ,t_{0}=0)$ for the reverse process is chosen. The reverse process that is used for inference starts with $\btX_{t_{N}} \sim \mathcal N_{\mathbb{C}}(\mathbf \bY, \sigma({t_{N}})^2 \mathbf I)$ and iteratively computes \cite{song2021sde}
\begin{align} \label{eq:em}
     \btX_{t_i} &= \btX_{{t_{i+1}}} -
    \left[
        \mathbf f(\mathbf \btX_{{t_{i+1}}}, \mathbf \bY) - g(t_i)^2  \mathbf s_\theta(\mathbf \btX_{{t_{i+1}}}, \mathbf \bY, t_i)
        \right] \Delta t(i) \notag \\ &
    + g(t_i)\sqrt{\Delta t(i)} \,\mathbf Z_{t_i}, 
\end{align}
where $ \Delta t(i) = {{t_{i+1}}} - {{t_{i}}}$ and $\mathbf Z_{t_i} \sim \mathcal N_{\mathbb{C}}(\mathbf 0, \mathbf I)$.
When fixing the step size $\Delta t(i)$, the reverse starting point $0 < \trsp \coloneqq t_{N} \leq T$ can be used for trading performance for computational speed. For optimal performance, one should set $\trsp = T$ to reduce the so-called prior mismatch \cite{lay202interspeech} between the terminating forward distribution and the initial reverse distribution. Using a smaller $\trsp$ reduces the number of iterations, but may also degrade the performance \cite{lay202interspeech}. The last iteration outputs $\btX_{t_0}$ approximating the clean speech signal.

\subsection{Brownian Bridge with Exponential Diffusion Coefficient (BBED)} \label{sec:bbed}
Recently, the BBED SDE \cite{lay202interspeech} has been shown to outperform existing SDEs \cite{paper, journal} for the task of SE in terms of several intrusive instrumental performance metrics. The BBED drift and diffusion coefficients are given by:
\begin{align} \label{eq:bb-drift}
    \mathbf f(\mathbf \bX_t, \mathbf \bY) &= \frac{\mathbf \bY-\mathbf \bX_t}{1-t},
\end{align}
and 
\begin{equation} \label{eq:bb-diffusion}
    g(t) = ck^t, ~~\text{where } c,k > 0
\end{equation}
for $0 \leq t \leq T < 1$, where $T$, as detailed in Section \ref{sec:sde}, is the terminating time for the forward SDE, which we set slightly smaller than $1$ to avoid division by zero in \eqref{eq:bb-drift}.
The closed-form solution for the variance of the perturbation kernel can be computed from \cite[(6.11)]{kara_and_shreve}:
\begin{align} \label{eq:bb:var}
   \hspace{-0.65em} \sigma(t)^2 &=  (1-t)c^2\left[(k^{2t}-1+t) + \log(k^{2k^2})(1-t)E \right], \\
    E &= \text{Ei}\left[2(t-1)\log(k)\right] - \text{Ei}\left[-2\log(k)\right],
\end{align}
where $\text{Ei}[\cdot]$ denotes the exponential integral function \cite{bender78:AMM}. The variance admits one peak between $0$ and $1$ and is zero for $t \in \{0, 1\}$. The mean can be computed from \cite[(6.10)]{kara_and_shreve} and linearly interpolates between the noisy mixture $\bY$ and the clean speech signal $\bX_0$:
\begin{equation}
\label{eq:bb:mean}
    \boldsymbol \mu(\mathbf \bX_0, \mathbf \bY, t) = \left(1- t\right)\mathbf \bX_0 + t \mathbf \bY
    \,.
\end{equation}

\section{Proposed Two-Stage Training} \label{sec:train}
To reduce the NFEs during inference, we now propose a two-stage training procedure. In the first training stage we train the score model on the \ac{dsm} loss \cite{songGenerativeModelingEstimating2020, journal, paper}. In the second training stage, we propose as a novelty to fine-tune the score model on a predictive loss which we denote in the following as \emph{correcting the reverse process} (CRP) loss.

\subsection{Denoising Score Matching (DSM)} \label{sec:train1}
Following Vincent \cite{vincent2011connection}, we fit the score model to the score of the perturbation kernel
\begin{equation} \label{eq:before_subsitution}
 \hspace*{-0.15em}\nabla_{\mathbf \bX_t} \log p_{0t}(\mathbf \bX_t | \mathbf \bX_0, \mathbf \bY) = -\frac{\mathbf Z}{\sigma(t)},
\end{equation}
where $\mathbf Z \sim \mathcal N_{\mathbb{C}}(\mathbf 0, \mathbf I)$. At each training step we proceed as follows: 1) we sample a training pair of clean speech data and noisy speech ($\mathbf \bX_0$, $\mathbf \bY$), 2) we sample $t$ uniformly from $[t_{\epsilon}, T]$, where $t_{\epsilon}>0$ is a small hyperparameter that assured numerical stability \cite{journal}, 3) we compute
$\mathbf \bX_t =  \boldsymbol\mu(\mathbf \bX_0,\mathbf \bY, t) + \sigma(t) \mathbf Z$. Finally, we obtain the \ac{dsm} loss, by minimizing the distance of the term on the right in \eqref{eq:before_subsitution} and the score model based on the $L_2$ norm:
\begin{equation}\label{eq:training-loss}
      \argmin_\theta \mathbb{E}_{t,(\mathbf \bX_0,\mathbf \bY), \mathbf Z, \mathbf \bX_t|(\mathbf \bX_0,\mathbf \bY)} \left[
        \norm{\mathbf s_\theta(\mathbf \bX_t, \mathbf \bY, t) + \frac{\mathbf Z}{\sigma(t)}}_2^2
    \right]
\end{equation}

\subsection{Correcting the Reverse Process (CRP)} \label{sec:train2}
By solving the reverse process as described in Section \ref{sec:solve_rsde}, we obtain an estimate $\btX_0$ of the clean speech signal $\bX_0$. 
This estimate, however, contains errors from several sources. First, there is the discretization error from the \ac{em} method. Second, there are errors caused by the score model over the discretization schedule. Moreover, there is the prior mismatch, as discussed in Section \ref{sec:solve_rsde}. This mismatch and the aforementioned error sources are not addressed using the \ac{dsm} loss in \eqref{eq:training-loss}.
This is because in the third step of computing the \ac{dsm} loss, we calculate $\bX_t$ by using the solution of the forward SDE given by \eqref{eq:perturbation-kernel}, but during inference, we compute the solution $\btX_t$ of the reverse SDE. After training on the \ac{dsm} loss, we therefore propose to retrain the score model to adapt to these errors and the prior mismatch. Specifically, we first fix a reverse starting point $\trsp$ and a discretization schedule $(\trsp = t_{N}, \dots ,t_{0}=0)$. Then, for each training pair of clean speech and noisy speech ($\mathbf \bX_0$, $\mathbf \bY$), we compute $\mathbf \btX_0$ based on \eqref{eq:em} by starting at $\trsp$ and taking steps according to the fixed discretization schedule, i.e. we specifically run the inference process.

In the second training stage, we then optimize on \ac{crp} loss 

\begin{equation} \label{eq:training-loss2}
     \argmin_\theta \mathbb{E}_{(\mathbf \bX_0,\mathbf \bY), \btX_0|(\mathbf \bX_0,\mathbf \bY)} \norm{\mathbf \btX_0 - \mathbf \bX_0}_2^2,
\end{equation}
by updating the weights based on the gradients of the last score model call $\mathbf s_\theta(\mathbf \btX_{t_1}, \mathbf \bY, t_1)$ used to obtain $\mathbf \btX_0$. This also means that when we iterate through \eqref{eq:em} the gradients of all other score model calls $\mathbf s_\theta(\mathbf \btX_{t_k}, \mathbf \bY, t_k)$, $k=N, \dots, 2$ are not used to update the weights of the neural network. The decision to exclusively update weights in the last score model call is driven by the fact that the accumulated discretization and score model errors can be corrected by the last score model call while keeping the cost of memory low.  %

Note that the idea of using a predictive loss for a diffusion model is conceptually similar to cold diffusion \cite{yen2023cold}. However, cold diffusion is not generative as it does not contain any randomness. %

\begin{figure*}
    \centering
    \includegraphics[scale=1]{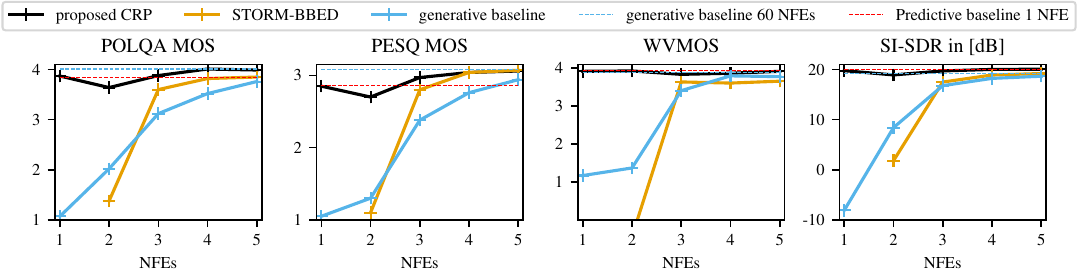}
    \caption{Performance results when trained and tested on WSj0-C3 of the proposed CRP, StoRM-BBED, predictive baseline and the generative baseline as a function of NFEs. In this work, the NFEs is the number of NCSN++ evaluations. The number of discretization steps $n_{\text{steps}}$ is chosen as described in Section \ref{sec:exp:crp}. We have that $n_{\text{steps}} = \text{NFE}$ for CRP and the generative baseline. For StoRM-BBED, we have $n_{\text{steps}} + 1= \text{NFE}$ (see Section \ref{sec:exp:baseline}). All solid lines use the same discretization schedule as described in Section \ref{sec:exp:crp}. Dotted blue line uses discretization schedule as in Section \ref{sec:bbed-para}.}
    \label{fig:1}
\end{figure*}

\begin{table}[t!] 
\centering
\begin{tabular}{|c|c|cc|}
\hline
Method & NFEs & POLQA & PESQ  \\
\hline
\hline
\multirow{1}{*}{generative baseline \cite{lay202interspeech}} & \multirow{1}{*}{60} & $3.61 \pm 0.65$ & $2.36 \pm 0.58$  \\
\hline
\hline
\multirow{1}{*}{predictive baseline} & \multirow{1}{*}{1}  &  $3.39 \pm 0.73$ & $2.14 \pm 0.57$  \\

\hline 
\hline
\multirow{2}{*}{proposed CRP} & \multirow{1}{*}{1}  & $3.47 \pm 0.71$ & $2.29 \pm 0.61$  \\
\cline{2-4}
& \multirow{1}{*}{5}  &  $\mathbf{3.65 \pm 0.65}$ & $\mathbf{2.55 \pm 0.59}$ \\
\hline
\end{tabular}
\caption{Speech enhancement results (mean and standard deviation) obtained for WSJ0-C3 when trained on VBD (mismatch case).}
\label{tab:res_vbd}
\end{table}

\section{Experimental setup} \label{sec:exp}
\vspace{-0.2cm}
\subsection{Metrics} \label{sec:exp:metric}
We evaluate the performance on perceptual metrics, wideband PESQ \cite{rixPerceptualEvaluationSpeech2001} and POLQA \cite{polqa2018}, on energy-based metrics SI-SDR \cite{lerouxSDRHalfbakedWell2018}. We also evaluate on a reference-free metric WVMOS using a neural network to predict MOS values.

The NFEs is the number of score model calls for producing the enhanced file. It is a measure of computational expenses during inference.

\vspace{-0.2cm}
\subsection{BBED parameterization} \label{sec:bbed-para}
A parameterization for the BBED SDE has already been grid-searched in \cite{lay202interspeech}. We therefore simply set $k=2.6$, $c = 0.51$ and $T=0.999$ as it is done in \cite{lay202interspeech}.

For inference, we employ the predictor-corrector scheme \cite{song2021sde}. The predictor is the \ac{em} method and the corrector is the annealed Langevin Dynamics method. We start the reverse process at $0.999$ and we use 30 discretization steps, resulting in 60 NFEs.

\subsection{Training on the \ac{dsm} loss} \label{sec:exp:dsm}
For the score model $s_\theta(\mathbf X_t, \mathbf Y, t)$, we employ the Noise Conditional Score Network (NCSN++) architecture (see \cite{journal, song2021sde} for more details). The first training stage follows mainly the setup from \cite{lay202interspeech}. More precisely, we first train the model on the \ac{dsm} loss function defined in \eqref{eq:training-loss} with a learning rate of $10^{-4}$ using the ADAM optimizer with a decay of $10^{-6}$. In addition, an exponential moving average (EMA) of the weights of NCSN++ is tracked with a decay of 0.999, to be used for sampling \cite{song2020improved}. Moreover, a lower bound for the diffusion step $t$ is set to be $t_{\epsilon}=0.03$ to avoid numerical instability (see \cite{journal, lay202interspeech} for details). Furthermore, we train for 200 epochs with a batch size of 16 and logged the averaged PESQ of 10 randomly selected files from the validation set during training and select the best-performing model for the second training stage.

\vspace{-0.2cm}
\subsection{Training on the \ac{crp} loss} \label{sec:exp:crp}
In the second training stage, we train on \eqref{eq:training-loss2} with the same ADAM and exponential moving average configurations as for the first training stage. We train on a batch size of 16 and train for only 10 epochs. We also validate during training on 10 randomly selected files from the validation set and chose the best-performing model accordingly for testing.

To train on the \ac{crp} loss in \eqref{eq:training-loss2}, we have to choose a discretization schedule. To this end, we set the reverse starting point $\trsp=0.5$. This choice is based on \cite{lay202interspeech}, where it has been observed that the BBED can perform for $\trsp = 0.5$ as well as for $\trsp = 0.999$ while reducing the \ac{nfe}. The discretization schedule is set as follows. We take $n_{\text{steps}} \in \{1, 2, 3, 4, 5\}$ discretizations steps as follows. The first $n_{\text{steps}}-1$ steps are uniformly from $\trsp$ to $\teps$ and the last step is taken from $\teps$ to 0. Moreover, when we take $n_\text{steps} = 1$, then we take only one step from $\trsp$ to $0$ directly. We do not use a corrector for training the \ac{crp} loss, therefore NFE = $n_{\text{steps}}$. In addition, we use the same discretization schedule for testing as for training.

\vspace{-0.2cm}
\subsection{Baselines} \label{sec:exp:baseline}
\textbf{Generative baseline:} We call the BBED SDE parametrized as in \ref{sec:bbed-para} and trained as described in Section \ref{sec:exp:dsm} the generative baseline. \\
\textbf{Predictive baseline:} We investigate the benefits of the generative BBED baseline over a predictive baseline by training the score model architecture on a predictive loss. We, therefore, train NCSN++ on the MSE loss between enhanced and the clean speech signal as it is done in the second training stage in \eqref{eq:training-loss2} and also use the same magnitude compressed STFT representation from Section \ref{sec:representation} for the input of the predictive NCSN++ baseline. Moreover, since there is no diffusion time embedding $t$ in the input for the predictive task, we removed the noise-conditioned layers. The modified architecture differs by less than 1 percent of its original size and therefore changes hardly the capacity of the network. Similar to training the \ac{dsm} loss, we use the ADAM optimizer with a learning rate of $10^{-4}$ and a decay of $10^{-6}$ for a maximum of 200 epochs. We train with a batch size of 16 and log the averaged PESQ of 10 randomly selected files from the validation set during training and select the best-performing model for testing. We do not employ EMA for tracking the weights.\\
\textbf{StoRM:} As the proposed method uses a generative loss and a predictive loss, we want to compare it against StoRM \cite{lemercier2022storm}, which is also using the exact same generative and predictive loss. StoRM has a predictive part and a generative part. In our setup, the predictive part uses the predictive baseline's NCSN++ architecture. The generative part of StoRM is due to the generative baseline. We call this modified version StoRM-BBED. For inference, we use the same discretization schedule as for the proposed method described in Section \ref{sec:exp:crp}. Note that the predictive part must always be executed once before solving the reverse SDE. Therefore, $\text{NFE}= n_{\text{steps}}+ 1$ for StoRM-BBED.

\vspace{-0.2cm}
\subsection{Datasets}
To analyze the generalization performance of the proposed method for SE, we cross-evaluate on two different datasets. Specifically, we train either on the VoiceBank-DEMAND dataset (VBD) or on the WSJ0-CHiME3 (WSJ0-C3) dataset, but we test only on WSJ0-C3. We choose to test on WSJ0-C3 over testing on VBD as the utterances of VBD have a short average length of 2.5 seconds, which is not ideal for evaluating on PESQ or POLQA \cite{PESQ}.

\noindent \textbf{VBD:}
 The publicly available VBD dataset \cite{valentini2016investigating} is commonly employed as a benchmark in single-channel SE tasks. Note that this dataset does not have a validation set. To ensure unbiased validation, we split the training data into two sets: training set and validation set. For validation, we specifically reserved the speakers ``p226'' and ``p287'' from the dataset. This has also been done in previous works \cite{journal, lay202interspeech}.

\noindent \textbf{WSJ0-C3:}
The WSJ0-C3 single-channel dataset that is also used in \cite{journal, lay202interspeech} mixes clean speech utterances from the Wall Street Journal dataset \cite{datasetWSJ0} to noise signals from the CHiME3 dataset \cite{barker2015third} with a uniformly sampled \ac{snr} between 0 and 20$\,$dB. The dataset is split into a train (12777 files), validation (1206 files) and test set (651 files). \\

\vspace{-0.3cm}
\section{Results} \label{sec:res}
We first show that the proposed method remains relatively stable when lowering the NFEs. Second, we show that the proposed method with a single NFE is not simply reduced to the predictive baseline, and third we show that the proposed method with only 5 NFEs performs as well as the generative baseline with 60 NFEs.
Moreover, we report results that the proposed CRP outperforms (unfolded) cold diffusion \cite{yen2023cold} by 0.5 in PESQ when both methods are trained and tested on VBD with one step (CRP achieves $2.96$ in PESQ).

First, from Fig. \ref{fig:1} we see the performance results of the proposed CRP, generative BBED baseline, predictive baseline and StoRM-BBED. In all plots of Fig. \ref{fig:1} we clearly observe the generative baseline (blue solid line) and StoRM-BBED (yellow solid line) also drop in performance when reducing the NFEs to few steps. For instance, both the generative baseline and StoRM-BBED achieve only around 1 on the MOS scale with NFE = 1 or NFE = 2 respectively. Note that a 1 on the MOS scale denotes "bad" quality according to ITU-T P.800. This is in contrast to the proposed CRP method, where performance remains relatively stable and achieves "good" (POLQA and WVMOS) or "fair" (PESQ) quality (see black solid line in Fig. \ref{fig:1}). 

Second, if only one NFE is applied, the performance of the proposed method seems to be reduced to the predictive baseline. For instance, in the POLQA plot, we report a value of 3.87 for the proposed method at $\text{NFE}=1$ and a very similar value of 3.85 for the predictive baseline. However, the advantage of the generative approach becomes visible in Tab. \ref{tab:res_vbd} where we show results for the mismatched case, i.e., we train on the VBD training set, but test on the WSj0-C3 test set. We observe that the proposed generative method outperforms the predictive baseline by 0.15 in PESQ also with only one NFE. These results imply that the better generalization to unseen data observed when comparing diffusion models against predictive models observed in \cite{journal, lemercierGenerativeVsDiscr} carries over to the proposed approach even when NFE = 1.

Third, we find that the proposed method achieves virtually the same performance results with 5 NFEs for the matched (see Fig. \ref{fig:1}, compare solid black line to dotted blue line) as the generative baseline with 60 NFEs and mismatched case (see Tab. \ref{tab:res_vbd}, compare first row to last row).

\vspace{-0.2cm}
\section{Conclusions}
In this work, we proposed to fine-tune a score-based diffusion model by optimizing a predictive loss to correct errors caused by solving the reverse process. We find that the proposed method remains relatively stable in performance when lowering the \acf{nfe}. In contrast, the pure generative approach and StoRM-BBED drastically reduce in performance when lowering the NFEs to a few or single reverse diffusion steps. Moreover, we showed that in contrast to a predictive baseline, the proposed method achieves an improved generalization to unseen data even with one diffusion step.

\bibliographystyle{ieeetr}
\bibliography{strings,refs}

\end{document}